# Combined Compute and Storage: Configurable Memristor Arrays to Accelerate Search

Yang Liu, Chris Dwyer, Alvin R. Lebeck

Duke University

## ABSTRACT

*Emerging technologies present opportunities for system designers to meet the challenges presented by competing trends of big data analytics and limitations on CMOS scaling. Specifically, memristors are an emerging high-density technology where the individual memristors can be used as storage or to perform computation. The voltage applied across a memristor determines its behavior (storage vs. compute), which enables a configurable memristor substrate that can embed computation with storage.*

*This paper explores accelerating point and range search queries as instances of the more general configurable combined compute and storage capabilities of memristor arrays. We first present MemCAM, a configurable memristor-based content addressable memory for the cases when fast, infrequent searches over large datasets are required. For frequent searches, memristor lifetime becomes a concern. To increase memristor array lifetime we introduce hybrid data structures that combine trees with MemCAM using conventional CMOS processor/cache hierarchies for the upper levels of the tree and configurable memristor technologies for lower levels.*

*We use SPICE to analyze energy consumption and access time of memristors and use analytic models to evaluate the performance of configurable hybrid data structures. The results show that with acceptable energy consumption our configurable hybrid data structures improve performance of search intensive applications and achieve lifetime in years or decades under continuous queries. Furthermore, the configurability of memristor arrays and the proposed data structures provide opportunities to tune the trade-off between performance and lifetime and the data structures can be easily adapted to future memristors or other technologies with improved endurance.*

## Categories and Subject Descriptors[1]
B.3.2 [**Hardware**]: Design Styles – *associative memories.* C.1 [**Computer Systems Organization**]: Procesor Architectures – *multiple data stream architectures, heterogeneous (hybrid) systems.* E.1 [**Data**] – *trees.*

## 1. INTRODUCTION
Workload and technology trends are significant driving forces behind computer systems design. Three significant current trends are large data sets, limits of CMOS power dissipation, and emerging technologies. First, the desire to query and analyze an increasingly large amount of data presents significant algorithm and systems challenges, e.g., [2, 6]. Second, the power dissipation limits of current CMOS packaging create an architectural trend toward the design of application accelerators that provide customized hardware for improving the performance of common workload scenarios [5, 12, 13, 34]. Third, scaling limits of CMOS motivate the need for alternative technologies to augment or supplant CMOS [3]. The confluence of these three trends presents an opportunity to explore new approaches that span traditional system abstraction boundaries from technology up through applications.

This paper explores memristors—an emerging high-density technology—where the individual memristors can be used either for non-volatile storage or to perform computation [8, 9, 18, 19, 29, 32, 35]. The voltage applied across a memristor determines its behavior (storage vs. compute), which enables configurable use of the memristor substrate to embed computation with storage. We propose using memristor arrays as a single combined compute/storage substrate that can be dynamically configured to provide customized computational support for big-data and other applications. In this paper, we focus on two types of search operations (point and range queries) as specific instances of the more general specialized accelerators. Search is an integral part of many applications including databases, machine learning, network routing, DNA sequencing; and recent research has explored methods for exploiting other new technologies for improving search [15] or database algorithms [7].

Memristors have the potential to provide higher capacity ($10^{12}/cm^2$) [32] than CMOS with switching times as low as 1ns an external array access times as low as 10ns [22, 29]. The memristive computation we explore is implication logic [4], which makes it possible to perform computation within the storage structure. Unfortunately, memristors have much lower endurance ($10^{10}$ write cycles [36]) than CMOS devices ($10^{16}$ write cycles for SRAM [11]) and in-storage computing further exacerbates the problem since each implication logic operation could be a memristor write. The challenge is to exploit the density and combined compute/storage aspect of memristors while maintaining acceptable lifetimes.

To meet the above challenges we first propose MemCAM, a configurable memristor-based content addressable memory (CAM). A search is performed by applying the same sequence of implication logic operations to each MemCAM cell in parallel. MemCAM can be used for either point or range queries by simply changing the allocation of memristors used for compute vs. storage and using a slightly different sequence of implication logic operations to perform greater than/less than comparisons instead of only equality. MemCAM is best suited for low query rates since its lifetime is only a few minutes under continuous queries. Standard wear leveling techniques are inadequate for MemCAM since all cells are accessed each query.

---
[1] Yang Liu is currently with Oracle, this work was performed while at Duke University.

To provide long lifetime under high query rates, we introduce configurable hybrid data structures that use both conventional CMOS processors/cache hierarchies and memristors for compute/storage. Our new data structures combine T-trees, B+-trees, and MemCAM to obtain a balance between search time and lifetime by exploiting a heterogeneous computing environment. The upper levels of the trees, accessed frequently, are implemented in software using conventional processors and caching methods and serve to distribute requests over the less frequently accessed remaining data—a technique we call algorithmic wear leveling. The memristor array and an associated programmable controller implements lower level tree traversal and/or MemCAM operations. These new data structures can be reconfigured to trade between performance and lifetime for a specific usage scenario and to adapt to future memristors with improved endurance.

The qualitative design space of memristor-based storage structures is shown in Figure 1. The lifetime of a memristor-based memory is the longest due to low write frequency and can be further improved by standard wear-leveling techniques. However, the search time of a memristor-based memory is the longest, and increases as data size increases. MemCAM has the shortest search time because all data items can be searched simultaneously but also has the shortest lifetime due to high write frequency. Wear leveling techniques cannot improve the lifetime of MemCAM because writes are already uniform. As long as endurance is limited for memristors, hybrid data structures are better choices because writes are distributed and occur less frequently per memristor.

To evaluate our designs we use SPICE to model an individual memristor and analyze energy consumption and performance. The results show that it is feasible to build a 1Gbit MemCAM with 1cm x 1cm area. For a K-bit search word, the energy consumption is $(0.44+0.82*\log_2(K))$ fJ/bit/search (for each data bit stored in MemCAM) and the search time is $16+20*\log_2(K))$ ns for MemCAM supporting both point and range queries, and the energy consumption is $(0.83+0.82*\log_2(K))$ fJ/bit/search and the search time is $(22+20*\log_2(K))$ ns for MemTCAM supporting both point and range queries. To evaluate the search performance and lifetime of the hybrid data structures we construct an analytic model, since it is impractical to simulate the large data sets required. We use 5nmx5nm memristors [22] ($10^{12}$ memristors per $cm^2$) instead of 50nmx50nm memristors ($10^{10}$ memristors per $cm^2$) so we can show the full potential of memristor-based storage structures to improve the performance of search operations. Our results show that hybrid storage structures can utilize range search abilities, achieve better performance than memory-based T-trees, and improve lifetime from minutes to longer than 60 years. Furthermore, TB+-tree-CAM, a hybrid memristor-based storage structure combining T-tree, B+-tree and CAM, manages to balance between performance and lifetime and can outperform other storage structures when taking both performance and lifetime into consideration.

We make three main contributions in this paper. First, this work takes the first step in exploring the combined compute/storage aspects of memristor arrays. Second, we propose configurable hybrid data structures to improve the performance and lifetime of search intensive applications. Finally, we provide configurability by using memristors as both storage and logic and by using both conventional CMOS processors/cache hierarchies and memristor technologies. Designers can choose to configure a memristor array as CAM, random access memory or hybrid CAM-memory to trade among power, capacity, performance and lifetime.

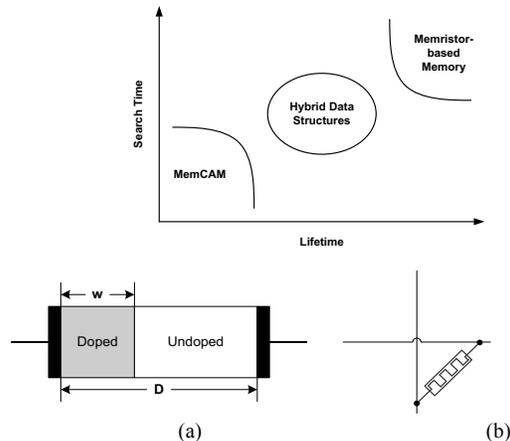

Figure 2: Memristor structure (a) and circuit notation (b)

We organize the remainder of this paper as follows: Section 2 introduces background knowledge. Section 3 summarizes our system overview. Section 4 describes in detail both cell design and match signal combination of MemCAM and the analysis of energy consumption and searching time. Section 5 proposes configurable hybrid memristor-based data structures and Section 6 evaluates the designs. Section 7 presents related work and Section 8 concludes.

## 2. Background
### 2.1 Memristors
The concept of a memristor was first predicted by Chua in 1971 [8] as the fourth fundamental circuit element and a physical model and prototype was recently presented by HP Labs [9]. A memristor is a non-volatile two-terminal nanoscale device that can switch states between 'on' (switch-closed) and 'off' (switch-open). A memristor array has ultra-high density (e.g. $10^{11}$ bits/$cm^2$ with a crossbar of approximately 17 nm half-pitch [17]) and could scale to 100 terabits/ $cm^2$ at 10nm feature sizes [32]. Figure 2 shows device schematic and cross bar circuit notation of a memristor. When a memristor is closed ($w \approx D$), it has low resistance and we consider it to represent logical value '1'; when a memristor is open ($w \approx 0$), it has high resistance and we consider it to represent logical value '0'. Recent proposals seek to utilize memristors to create novel nanostores for use in providing high-capacity nonvolatile memory for big-data workloads [28]. Our work seeks to complement that work by exploiting the additional capability of memristor arrays to perform computation.

The natural logical operation to compute with memristors is material implication $p \rightarrow q$ [18]. Figure 3 shows two memristors used to perform implication logic. The voltage applied on memristor $p$, $V_{COND}$, is a reading voltage, which does not change the state of $p$. The voltage applied on memristor $q$, $V_{SET}$, is a writing voltage that may change the state of $q$ depending on the initial states of both $p$ and $q$. $R_G$ is a resistance chosen between the 'on' state resistance and the 'off' state resistance. From the truth table in Figure 3 we can see that if we initialize q to be 0, the two memristors perform a NOT operation, q = ¬p. As we show later, other more complex operations are possible and can be performed in parallel. Although we focus on memristors in this paper, our techniques are applicable to any technology with similar properties.

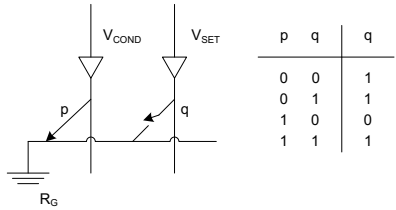

**Figure 3: Memristor Implication Logic for q = ¬p.**

## 2.2 Alternative Implementations

Associative lookup can be implemented in software (e.g., hash tables, balanced trees, etc.) and some languages (e.g., perl, java, python, etc.) provide direct support for data structures that expose the associative lookup interface (i.e., maps, associative arrays). Software implementations work very well for small data sets and applications that are latency and bandwidth tolerant. For applications with large data sets, software associative lookup implemented on commodity hardware can incur significant delays when the data set is too large to fit in conventional CMOS physical memory and long latency disk accesses are required. The high-density of emerging memories provides the opportunity to provide much larger physical memory reducing the need for external disk access in many applications. Furthermore, software implementations generally require a logarithmic number of memory accesses (e.g., balanced tree access). For applications that require sustained high throughput, this logarithmic number of accesses may be unacceptable even for data sets that can fit into memory. Hash tables may reduce the number of accesses to O(1) but at the expense of underutilized memory capacity since collisions must be avoided. This wasted memory capacity may be unacceptable for many applications.

An alternative to software associative lookup is to provide direct hardware support (specialization) in the form of content addressable memory (CAM). These specialized memories provide additional circuitry to simultaneously compare the content of each location to a provided key and returning either the data associated with the key or a set of addresses for entries with matching keys. This additional circuitry introduces overhead in terms of power consumption and access time. These overheads can limit the capacity of CAMs implemented in CMOS technology. Additional delays could be incurred since in many applications, the address of a matching entry is used to access other storage such as DRAM or disk. The capacity of CMOS-based CAMs may also be limited by the rate of scaling.

Memristors and other emerging high-density memories (e.g., STTRAM) could be used to create dedicated CAMs [11, 15]. However, combining CMOS transistors with memristors unnecessarily limits density and increases manufacturing difficulty since the CAM cell size is determined by CMOS device sizes rather than memristor device sizes. Alternatively, a specialized design using only memristors could be used to create a CAM [30]. Although these techniques could increase CAM capacity, traditional hardware CAMs are limited to equality comparisons and would incur significant capacity reductions to provide support for even slightly more complex operations (e.g., range query). Therefore, we seek to complement the capacity advantages of an all memristor design with the flexibility of configurable computation allowing designs to be tailored to individual application requirements.

Many applications perform more than just a simple comparison and thus can benefit from more general computational ability in the accelerator. High-density resistive memory can also be used similar to FPGAs by configuring lookup tables (LUTs) to create specified circuits [16]. The work in this paper differs in that we seek to exploit the ability of memristor's to perform implication logic (thus computation) in a programmable manner by controlling the voltages across memristors. LUT-based computing is ideal for technologies where write latency/power is much greater than read latency/power. We expect memristor write and read characteristics to be roughly equal and may be as low as 10ns [22, 29]. Nonetheless, exploring the tradeoffs between LUT-based computing and sequencing implication logic steps is an interesting avenue to explore in future work.

## 3. System Overview

Our overall system design is shown in Figure 4. Although this structure places the memristor array on the physical memory bus along with conventional DRAM modules, it is possible to also utilize a 3D stacked fabrication process similar to that advocated for creating nanostores [28, 32]. Regardless of the specific packaging approach, we envision a memristor array that resides in the system's physical address space.

The memristor subsystem is composed of a memristor array and a programmable controller. The processor communicates with the memristor array controller using memory-mapped operations. The controller is responsible for applying appropriate voltages to perform read/write or implication logic operations using the memristor array. Read/write operations are 'external' operations since peripheral CMOS circuitry is required to decode the address, evaluate the data read out (for reads) and decide the applied voltages based on the data to write (for writes). In contrast, implication logic operations are 'internal' operations on data already stored in memristors and the results are generated and stored in memristors without being read out externally. Therefore, external accesses will take much longer than the internal implication logic steps. Applying voltages to perform a series of implication logic steps in sequence performs computation. Note that this design does not cascade memristors to create combinational circuits, in contrast to conventional CMOS transistors. However, parallelism can be exploited by using many memristors to perform multiple implication logic operations per step.

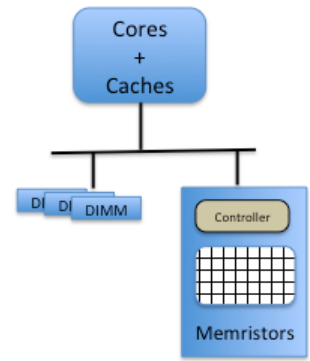

**Figure 4: System Overview of Configurable Memristor Array**

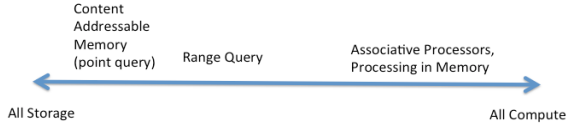

**Figure 5: Spectrum of Configurable Memristor-based Computing**

We assume a programmable memristor array controller where the program specifies the sequence of voltages to apply to the memristor array. Partitioning of memristors between storage and computation is entirely under software control since it is the voltages that determine compute vs. storage. We assume the controller can always perform read/write operations to any portion of the memristor array, even the memristors used for computation. Configuration/specialization occurs by specifying a particular program for the controller to execute that augments the traditional read/write memory behavior. Unfortunately, there may not be arbitrary flexibility in mapping computation onto the memristor array while still providing high performance.

To achieve high density, crossbar arrays are used in the memristor array and thus voltages are applied to entire rows and entire columns. Although it is possible set individual memristor voltages using this two-dimensional array, the rate of computation may be very slow. Instead, the mapping of computation onto the memristor array should exploit the two-dimensional structure such that many memristors can share a single voltage setting and thus achieve parallel operation. In this work we perform manual configuration/mapping of computation onto the memristor array, but automated mapping is an interesting avenue of future work.

The configurability of memristor arrays creates a spectrum of potential designs. As shown in Figure 5, on one end of the spectrum the memristor array is configured to provide only storage and can be used as nonvolatile memory while at the other extreme is pure computation. In between these end points is a diverse set of options for providing customized application accelerators. In this paper we focus on search operations and leave exploration of more sophisticated acceleration as future work.

## 4. MemCAM: memristor-based CAM

This section presents our memristor CAM design (MemCAM). We begin with a description of a single MemCAM cell. We focus on CAM cell design and match signal combination. We assume peripheral circuitry required to write into and read from the memristor array similar to that proposed elsewhere [35]. We designed both CAM and TCAM using memristors with similar comparison and match signal combination processes. For brevity, we only present the details of the memristor TCAM design that supports both point and range query, we continue to use the generic term CAM to refer to this implementation. If only CAM operations are required then a slightly different design could be configured that uses fewer memristors per entry.

### 4.1 MemCAM cell design

Figure 6 shows how memristors in an array are organized to form rows of MemCAM entries. Each row contains multiple entries

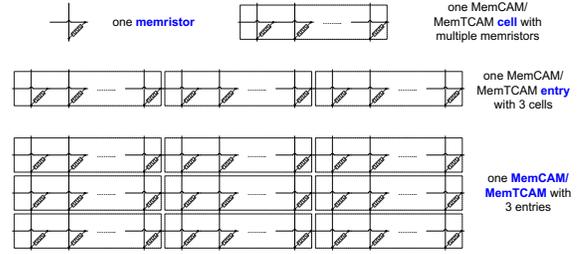

**Figure 6: MemCAM/TCAM Organization**

**Table 1: Values and meanings of cell match signals ($M_3$ & $M_4$) based on stored Data and key bits.**

| $D_1$ | $D_0$ | K | $M_3 = D_1 \wedge \neg D_0 \wedge K$ | $M_4 = D_1 \wedge D_0 \wedge \neg K$ | |
|---|---|---|---|---|---|
| 0 | 0 | 0 | 0 | 0 | $D == K$ |
| 0 | 1 | 1 | 0 | 0 | |
| 0 | 1 | 0 | 0 | 1 | $D > K$ |
| 0 | 0 | 1 | 1 | 0 | $D < K$ |
| 1 | X | X | 0 | 0 | $D == K$ |

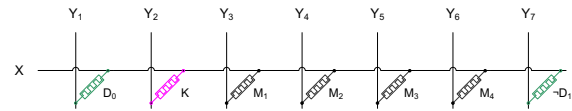

**Figure 7: MemTCAM cell design: each box is a memristor at a junction of the crossbar array.**

(for simplicity we show only one entry per row), each entry contains multiple cells, and each cell is comprised of multiple individual memristors. Figure 7 shows a MemCAM cell that can be used for both point and range queries. $D_0$ and $\neg D_1$ are two memristors used to store two bits representing the data bit, and K is the memristor used to store the input key bit. We store $\neg D_1$ instead of $D_1$ in order to save one step during the comparison process. $M_1$ to $M_4$ are memristors used to perform comparison and store match signals. $M_1$ and $M_2$ are used to store $\neg D_0$ and $\neg K$ first. K and $M_2$ are then used to compute $D_0 \vee \neg K$ and the result is stored in $M_2$, and $D_0$ and K are used to compute $\neg D_0 \vee K$ and the result is stored in K. $M_4$ is then used to store the value of $D_1$ and combined with the values of $M_2$ and K. Finally, $M_3$ and $M_4$ are cleared and used to store the match signal for the MemTCAM cell.

Table 1 shows the values and meanings of cell match signals based on the values of D ($D_1 D_0$) and K. The comparison process includes eleven steps. Table 2 shows voltages applied to the control lines, X and $Y_1$-$Y_7$, in a CAM cell at each step. The difference between voltages applied on two control lines connected to a memristor is the voltage across the memristor. $V_{CLEAR}$ is the voltage required to switch a memristor to its 'off' state. Table 2 also shows the states of $M_1$ through $M_4$ at each step. During the comparison process, the states of $D_0$ and $\neg D_1$ are not changed so their states are not shown in Table 2.

### 4.2 Match signal combination

**Table 2: Memristor States and Applied voltages at Each Step of Comparison for Point and Range Query with TCAM**

($V_{CO} = V_{COND}$, $V_S = V_{SET}$, $V_{CL}=V_{CLEAR}$)

| | K | $M_1$ | $M_2$ | $M_3$ | $M_4$ | $Y_1$ | $Y_2$ | $Y_3$ | $Y_4$ | $Y_5$ | $Y_6$ | $Y_7$ |
|---|---|---|---|---|---|---|---|---|---|---|---|---|
| Step 1 | $K$ | 0 | 0 | 0 | 0 | 0 | 0 | $V_{CL}$ | $V_{CL}$ | $V_{CL}$ | $V_{CL}$ | 0 |
| Step 2 | $K$ | $\neg D_0$ | 0 | 0 | 0 | $V_{CO}$ | 0 | $V_S$ | 0 | 0 | 0 | 0 |
| Step 3 | $K$ | $\neg D_0$ | $\neg K$ | 0 | 0 | 0 | $V_{CO}$ | 0 | $V_S$ | 0 | 0 | 0 |
| Step 4 | $K$ | $\neg D_0$ | $D_0 \vee \neg K$ | 0 | 0 | 0 | 0 | $V_{CO}$ | $V_S$ | | 0 | 0 |
| Step 5 | $\neg D_0 \vee K$ | $\neg D_0$ | $D_0 \vee \neg K$ | 0 | 0 | $V_{CO}$ | $V_S$ | 0 | 0 | 0 | 0 | 0 |
| Step 6 | $\neg D_0 \vee K$ | $\neg D_0$ | $D_0 \vee \neg K$ | 0 | $D_1$ | 0 | 0 | 0 | 0 | 0 | $V_S$ | $V_C$ |
| Step 7 | $\neg D_0 \vee K$ | $\neg D_0$ | $\neg D_1 \vee D_0 \vee \neg K$ | 0 | $D_1$ | 0 | 0 | 0 | $V_S$ | 0 | $V_{CO}$ | 0 |
| Step 8 | $\neg D_1 \vee \neg D_0 \vee K$ | $\neg D_0$ | $\neg D_1 \vee D_0 \vee \neg K$ | 0 | $D_1$ | 0 | $V_S$ | 0 | 0 | 0 | $V_{CO}$ | 0 |
| Step 9 | $\neg D_1 \vee \neg D_0 \vee K$ | $\neg D_0$ | $\neg D_1 \vee D_0 \vee \neg K$ | 0 | 0 | 0 | 0 | 0 | 0 | 0 | $V_{CL}$ | 0 |
| Step 10 | $\neg D_1 \vee \neg D_0 \vee K$ | $\neg D_0$ | $\neg D_1 \vee D_0 \vee \neg K$ | $D_1 \wedge \neg D_0 \wedge K$ | 0 | 0 | 0 | 0 | $V_{CO}$ | $V_S$ | 0 | 0 |
| Step 11 | $\neg D_1 \vee \neg D_0 \vee K$ | $\neg D_0$ | $\neg D_1 \vee D_0 \vee \neg K$ | $D_1 \wedge \neg D_0 \wedge K$ | $D_1 \wedge D_0 \wedge \neg K$ | 0 | $V_{CO}$ | 0 | 0 | 0 | $V_{SET}$ | 0 |

After each CAM cell finishes the eleven-step comparison and generates its cell match signal (CMS), we need to combine the match signals from all cells in an entry to generate the entry match signal (EMS).

We have:

$$EMS_i = Combination of (CMS_{i,0}, CMS_{i,1}, CMS_{i,2},..., CMS_{i,n-1})$$

in which $EMS_i$ is the match signal of the $i^{th}$ entry in the CAM, $CMS_{i,j}$ is the match signal of the $j^{th}$ cell in the $i^{th}$ entry, and $n$ is the number of cells in an entry, which is also the number of bits in the key word.

We assume $n$ to be power of two here and use recursive doubling [31] to combine match signals. For each Entry $i$, we first combine every CMS pair, $CMS_{i,2j}$ and $CMS_{i,2j+1}$ simultaneously and store the result in the memristor used to store $CMS_{i,2j+1}$. We then combine every CMS pair $CMS_{i,4j}$ and $CMS_{i,4j+2}$ similarly. $EMS_i$ is in the memristor used to store $CMS_{i,n-1}$ after $\log_2(n)$ rounds. Each round of match signal combination includes ten steps. We use six memristors from two adjacent cells, including four memristors already storing the CMSs, to combine two CMSs from two cells.

## 4.3 Discussion

Using memristors as both memory and logic provides not only high density but also configurability. Consider three alternatives: 1) all memory, 2) all CAM or 3) partitioned memory+CAM. Furthermore, for any CAM portion, we can configure different number of entries with different key sizes, including very large keys (e.g., character strings). Specific configurations can be based on application requirements.

However, one major disadvantage of MemCAM is that memristors have much lower endurance ($10^{10}$ write cycles) than SRAM ($10^{16}$ write cycles). The lifetime of MemCAM is only a few minutes under continuous search operations. Unfortunately, memCAM's lifetime cannot be improved by standard wear leveling techniques since all the cells are accessed simultaneously every cycle. To solve this problem, we need to design storage structures that reduce the average write frequency per cell.

## 5. Configurable Hybrid Data Structures

This section presents several novel hybrid data structures for point and range queries that are designed to take advantage of the in-place compute capabilities of memristors while alleviating the wear-out limitations. They key insight behind our approach is to design data structures that naturally distribute operations over the memristor array.

### 5.1 Overview

We can reduce the average write frequency by utilizing the configurability of a memristor array. We can divide a memristor array into multiple partitions with each partition having the same capacity and configure one partition as CAM and the other partitions as memory. We can then 'rotate' the CAM partitions within the memristor array to achieve the benefit of wear leveling.

The improvement of lifetime by using the hybrid memristor-based CAM-memory design is approximately proportional to the number of partitions. However, this design requires a large memristor array to obtain acceptable lifetime of a small MemCAM. For example, to achieve one month-lifetime for 1MB of MemCAM with continuous search operations requires a 35GB memristor array even if there are no writes to the memory partitions. With the improvement of memristor endurance in the future, this design may become more efficient, but currently the high storage overhead of the memristor-based CAM-memristor design makes it not practical.

We can also reduce write frequency by designing a hierarchical storage structure. We can use a CMOS-based CAM as a buffer of MemCAM. We store hot data (data searched more frequently) in CMOS CAM buffer and store cold data in MemCAM. The search frequency of MemCAM is reduced and so is the write frequency. The improvement of lifetime by using the hybrid CMOS-memristor-based CAM design is dependent on the capacities of both CAMs and the access frequencies of both hot and cold data.

Hot data has to be accessed $4 \times 10^5$ times more frequently than cold data in order to achieve a one-year lifetime, which is unlikely for many applications and limits the application area of this design.

Partitioning a memristor-based storage structure or adding a CMOS-based buffer alone cannot efficiently reduce write frequency. Thus we combine the two methods and propose a series of configurable hybrid data structures to utilize the computation ability of memristors and provide 'algorithmic' wear-leveling to improve lifetime. We start with a logical tree structure and divide it into two parts, the upper levels (the levels near the root) and the lower levels (the levels near the leaves). We then implement the two parts with different data structures and technologies. The upper levels can be implemented as a hash table or a T-tree and are stored in a CMOS-based storage structure (e.g., cache), and the lower levels can be implemented as a CAM or several $B^+$-trees and are stored in a memristor-based storage structure. The main idea is to direct search through the upper-level implementation so only one part of the memristor-based storage structure is accessed per search (one or two partitions of CAM, or one or two $B^+$-trees). The improvement of lifetime is proportional to the number of CAM partitions or $B^+$-trees when the accesses are uniformly distributed. When the accesses are not uniformly distributed, we can apply wear-leveling techniques to improve lifetime.

We decide the implementations of the two parts of the logical tree based on whether we can efficiently generate a hash function that is both uniform and order-preserving. A hash function is uniform if it maps the expected input as evenly as possible over its output range, and a hash function F is order-preserving if for inputs $k_1$ and $k_2$, $k_1 < k_2$ implies $F(k_1) < F(k_2)$. The properties of hash functions, together with the implementations, decide the functionality of the data structure – whether it can support range search or not.

When we can efficiently generate a hash function that is both uniform and order-preserving, we implement the upper levels as a hash table and the lower levels as a CAM (Hash-CAM). When we can efficiently generate a hash function which is only uniform but not order-preserving, we can still implement the logical tree as Hash–CAM but can only perform point search, which means that the comparison process can only decide that whether an entry is equal to the input key or not. If we also want to perform range search, in which we want to know whether an entry is greater than, or less than, or equal to the input key, we have to implement the upper levels as a data structure with sorted data instead of a hash table. We choose to implement the upper levels as a T-tree in this case (T-tree-CAM). Based on T-tree-CAM, we propose $TB^+$-tree and $TB^+$-tree-CAM to provide more configurability so we can further improve lifetime.

## 5.2 Hash-CAM

A Hash-CAM is a hybrid hash table and CAM data structure used to implement a logical tree. The hash table is used to implement the $i^{th}$ level of the tree with one node stored in one hash table entry. The CAM is divided into multiple partitions and one partition is linked with one hash table entry as shown in Figure 8. The hash table is used to store keys to direct search into one part of the CAM and the CAM is used to store all the records.

For point search, the input key goes through the hash function and the search is directed to one CAM partition. The corresponding CAM partition is searched with the process described in Section 4 and the matched results are read out based on entry match signals. For range search, the two input bound keys go through the hash function and the search is directed to two CAM partitions (bound

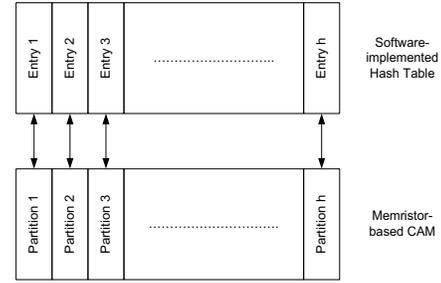

**Figure 8: HASH CAM**

CAM partitions). The two CAM partitions perform comparisons and output records with keys within the given range and any records in the partitions between the two bound CAM partitions.

From the search process we can see that at most two CAM partitions perform computations per search. As a result, the improvement of lifetime is proportional to the number of CAM partitions (which is also the number of hash table entries) when searches are uniformly distributed among all CAM partitions or when searches are not uniformly distributes and wear-leveling techniques are applied to rotate data among CAM partitions.

## 5.3 T-tree-CAM

If we can only efficiently generate hash functions that are only uniform but not order-preserving, Hash-CAM can only support point search but not range search because records within a range may be distributed among all CAM partitions. In order to support range search, we replace the hash table with a T-tree to implement the upper levels of the logical tree.

A T-tree is a data structure evolving from AVL trees and B-trees and mainly used in main-memory databases [20]. Figure 9 shows a T-tree node (T-node). It has a binary search nature similar to an AVL tree because it is a binary tree, and it has good update and storage characteristics similar to a B-tree because there are multiple elements per node. Compared with AVL trees, a T-tree requires fewer rotations upon delete and insert operations for rebalancing because of intra-node data movement.

We implement the upper levels of the logical tree with a T-tree to

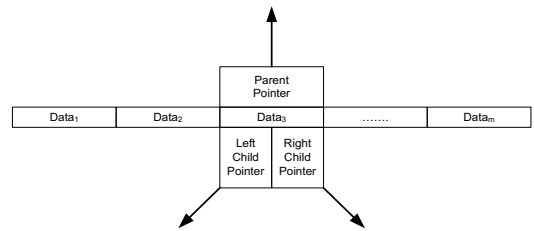

**Figure 9: A T-tree Node (T-Node)**

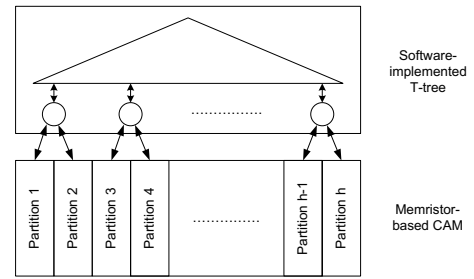

**Figure 10: A T-tree CAM**

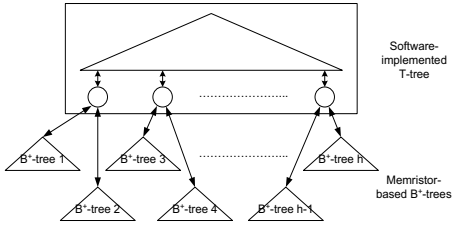

**Figure 11: TB+-tree**

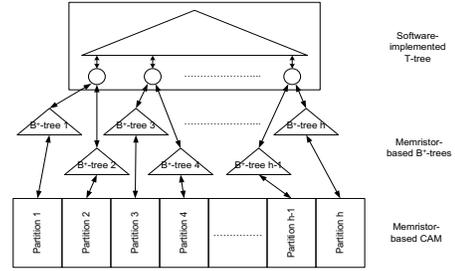

**Figure 12: TB+-tree-CAM**

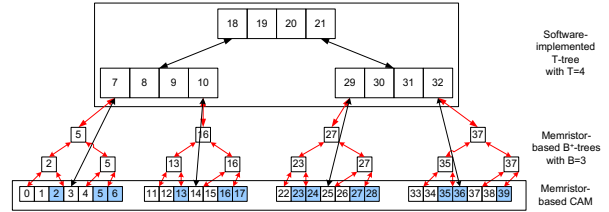

a) Faster Search / Shorter Lifetime (4 CAM Partitions)

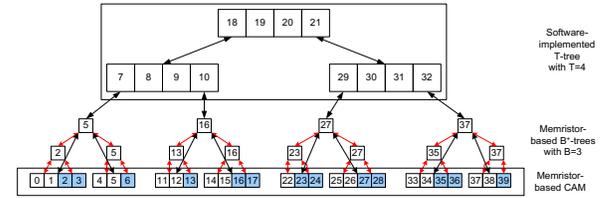

b) Slower Search / Longer Lifetime (8 CAM Partitions)

**Figure 13: TB+tree-CAM Tuning Options w/ Support for Insert & Delete**

preserve the orders to support range search. The lower levels are implemented with a CAM. The CAM is divided into multiple partitions and one partition is linked with one node in the lowest level of the T-tree as shown in Figure 10. Both point search and range search in a T-tree-CAM are similar to a Hash-CAM. The only difference is that the input keys go through a T-tree instead of a hash function. As a result, the improvement of lifetime is also proportional to the number of CAM partitions (which is also the number of nodes at the lowest level of the T-tree) when wear-leveling techniques are applied.

## 5.4 TB+-tree

T-tree-CAM does not require a uniform and order-preserving hash function to improve lifetime of memristor-based storage. However, the lifetime improvement is limited by the capacity of CMOS-based storage. In order to solve this problem, we propose a new hybrid data structure—a TB$^+$-tree. A TB$^+$-tree is a combination of T-tree and B$^+$-tree. The upper levels of a logical tree are implemented by a T-tree and stored in CMOS-based storage and the lower levels are implemented by a forest of B$^+$-trees, stored in memristor-based storage, and traversed within the memristor array using the memristor controller. Each B+ tree is linked with a node at the lowest level of the T-tree as shown in Figure 11. For point search, we go through one path in one B$^+$-tree to the leaf. For range search, we go through two paths in one or two B$^+$-trees to the leaves. Only a part of at most two B$^+$-trees, not two complete B$^+$-trees, perform computations per search. We can obtain lower average write frequency (thus longer lifetime) compared with T-tree-CAM. We can also achieve more configurability based on changing the order of B$^+$-tree.

We implement lower levels using B$^+$-tree instead of T-tree because B$^+$-tree is shallower than T-tree, which reduces the average time required to perform search/delete/insert operations. B$^+$-tree is not efficient for traditional main-memory databases because binary search is required to search within a sorted node and linear search is required to search within an unsorted node [7] which significantly increases search time. However, the intra node search time can be improved by using the memristor array to perform comparisons between the input key and all the keys stored in a B$^+$-tree node simultaneously. As a result, we can fully utilize the benefits of unsorted nodes to reduce write frequency.

## 5.5 TB$^+$-tree-CAM

Both T-tree-CAM and TB$^+$-tree have advantages and disadvantages. T-tree-CAM has shorter search time but limited lifetime. TB$^+$-tree have longer lifetime but also longer search time. In order to balance performance and lifetime, we propose another configurable hybrid data structure in between, a TB$^+$-tree-CAM as shown in Figure 12. In TB+-tree-CAM, we group leaf nodes of one subtree in one B$^+$-tree and align them continuously in the memristor array so we can perform CAM search operations described in Section 5.

The TB$^+$-tree-CAM is the most general data structure and the previous tree-based structures can be viewed as degenerate cases that enable tuning an application to trade off performance (search latency) vs. lifetime. Figure 13 shows two options for tuning while maintaining support for insert/delete operations. Search operations follow black arrows and insert/delete operations follow gray arrows. In general, the root of the CAM allocated subtree can be any node of one B+-tree. If the root of the subtree is the root of the B$^+$-tree, the TB$^+$-tree-CAM becomes a T-tree-CAM (Figure 13a). If the root of the subtree is one leaf node, TB$^+$-tree-CAM becomes TB$^+$-tree. If the root of the subtree is an internal node, TB$^+$-tree-CAM becomes a data structure in between with moderate search time and lifetime (Figure 13b).

## 5.6 Discussion

We propose four hybrid data structures in this section. All the designs are based on a logical tree divided into two parts, the upper levels and the lower levels. The main idea is to partition the lower levels and for every search/insert/delete operations, direct access to one or two of the partitions through the upper levels. Since at most two partitions are accessed per operation, the write frequency is reduced for the same number of operation, which leads to lifetime improvement proportional to the number of partitions. We can decrease the number of partitions by decreasing the number of upper levels (an extreme case is MemCAM, in which the lower levels are implemented with a CAM and the number of partitions is 1) or increase the number of partitions by increasing the number of upper levels. Users/Designers can choose different numbers of partitions to trade between performance and lifetime based on the requirements of different

**Table 3: Analytic Model Parameters**

| | |
|---|---|
| AvgTime$_T$ | Average record search time of the T-tree |
| AvgTime$_{TB}$ | Average record search time of the TB$^+$-tree |
| Level$_U$ | Number of upper levels |
| Level$_{LT}$ | Number of lower levels in the T-tree |
| Level$_{LTB}$ | Number of lower levels in the TB$^+$-tree |
| Level$_{LTB'}$ | Number of lower levels in the TB$^+$-tree-CAM |
| NodeTime$_U$ | Time to search a node at upper levels |
| NodeTime$_{LT}$ | Time to search a node at lower levels in the T-tree |
| NodeTime$_{LTB}$ | Time to search a node at lower levels in the TB$^+$-tree |
| N$_T$ | Total number of nodes storing records in the T-tree |
| N$_{TB}$ | Total number of nodes storing records in the TB$^+$-tree |
| N$_R$ | Total number of records |
| T | Order of the T-tree |
| B | Order of the B+-tree |

applications or when the endurance of memristors are improved by future research.

## 6. Evaluation

We develop an analytic model to evaluate and compare the average record search time of six data structures, a CMOS-based T-tree, a memristor-based T-tree, a hybrid Hash-CAM, a hybrid T-tree-CAM, a hybrid TB$^+$-tree and a hybrid TB$^+$-tree-CAM. All six data structures have the upper levels stored in a CMOS-based cache. The CMOS-based T-tree has the lower levels stored in DRAM. The memristor-based T-tree has the lower levels stored in a memristor memory and uses conventional loads and stores to traverse the tree. The four hybrid data structures store the lower levels in a memristor array with combined compute and storage and can leverage the internal controller to traverse the trees.

### 6.1 Energy and Access Feasibility Study

We first evaluate the energy consumption and search time of MemCAM and then evaluate hybrid storage structures based on MemCAM performance. Although we anticipate $10^{11}$ memristors/cm$^2$ if we build the memristor array on 17-nm-wide nanowires [17] and $10^{12}$ memristors/cm$^2$ with 5nm-scale memristors [1], we evaluate energy consumption and searching time of MemCAM based on a conservative design, a memristor array built on 50-nm-wide nanowires [4] with 50nm x 50nm x 10nm memristors. Memristor density of the evaluated array is $10^{10}$ memristors/cm$^2$ and cell density is $10^9$ cells/cm$^2$, which is 100 times denser than CMOS-based CAM.

We use a simplified SPICE model proposed by Mahvash and Parker [25] to simulate switching time and power consumption of a single memristor. The simulation results are then used to calculate the energy consumption and search time of MemCAM. The energy consumption and search time both depend on step time (time required to perform a step of operation). Step time depends on both switching time and RC delay, which can overlap because switching starts as soon as the voltage across a memristor goes beyond a threshold voltage (the lowest voltage that can switch a memristor). Based on the RC delay of 35 nm Cu-Low κ technology (250 ps for a 1 mm line [1]), methods such as repeater insertion are required to obtain a < 200ps RC delay for a 1-cm-long 50-nm-wide line and we can then obtain a 2-ns step time. The final results show that it is feasible to build a 1Gbit MemCAM with 1cm x 1cm area. With 50nmx50nm memristors and K-bit keywords, for MemCAM supporting both point and range queries, the energy consumption is $(0.44+0.82*\log_2(K))$ fJ/bit/search (for each data bit stored in MemCAM) and the search time is $(16+20* \log_2(K))$ ns, and for MemTCAM supporting both point and range queries, the energy consumption is $(0.83+0.82* \log_2(K))$ fJ/bit/search and the search time is $(22+20* \log_2(K))$ ns.

Based on the power consumption of a single memristor, we also estimate the power density of MemCAM. The power density of MemCAM is determined based on the power consumed by both memristors and wires. Previous studies show that wires consume up to 80% of the power [4]. However, in this experiment, the number of memristors and the number of wires are similar while in a 1cm x 1cm MemCAM there are $10^{10}$ memristors but only 2 x $10^5$ wires. As a result, the wire power percentage in MemCAM should be much lower. Furthermore, wire power density can be reduced by methods that could dramatically reduce the wire resistance and capacitance [4] since interconnect power is proportional to the wire capacitance [2,36]. We conservatively assume that wires consume 50% of the total power, which leads to a total power density of approximate 55W/cm$^2$ for MemCAM/MemTCAM supporting only point query and 80W/cm$^2$ for MemCAM/MemTCAM supporting both point and range queries. We expect similar power density as memristor feature size scales down, reaching the $10^{12}$ memristors/cm$^2$ density. The reason is that there is a linear relationship between the number of memristors per unit area and the memristor resistance and the power density depends on the ratio of the number of memristors per unit area to the memristor resistance. However, we must wait for experimental demonstrations of high-density memristor arrays to further analyze the power dissipation.

### 6.2 Analytic Model

Table 3 shows the parameters we use to develop our analytic model. We define *record nodes* as nodes containing pointers to the records. For the T-tree, all the nodes are record nodes. For the TB$^+$-tree, all the nodes at upper levels and all the leaves are record nodes. Given the number of records $N_R$, the order of the T-tree $T$ (which means that there are at most $T$ records in a T-node), the order of the B$^+$-tree $B$ (which means that each internal node in the B$^+$-tree has at least $\lceil B/2 \rceil$ children and at most $B$ children), and the number of upper levels *Level$_U$*, with the assumption of 100% node utilization (which means that every node has the maximum number of records) the total number of record nodes in the T-tree is $N_T = \frac{N_R}{T}$ and and the TB$^+$-tree the number of record nodes is $N_{TB} = 2^{Level_U} - 1 + \frac{N_R - (2^{Level_U} - 1) \times T}{B}$ .

The total number of record nodes in the T-tree ($N_T$) is calculated by dividing the number of record (N$_R$) by the order of the T-tree (*T*). The total number of record nodes in the TB+-tree ($N_{TB}$) is the sum of the number of record nodes at the upper levels and the lower levels, and the number of record nodes at the lower levels is calculated by dividing the number of record nodes at the lower levels by the order of the B$^+$-tree (*B*).

We then determine the number of lower levels in the T-tree ($Level_{LT}$) and the TB$^+$-tree ($Level_{LTB}$). The number of lower levels in the T-tree is the difference between the total number of levels in the T-tree and the given number of upper levels: $Level_{LT} = \left\lceil \log_2 \frac{N_R}{T} \right\rceil - Level_U$. We assume that all the B$^+$-trees in the TB$^+$-tree have the same depth and the number of lower levels in the TB$^+$-tree is equal to that depth, which is the logarithm of the number of record nodes in a B$^+$-tree to based $B$, which produces $Level_{LTB} = \left\lceil \log_B \left[ \frac{N_R - (2^{Level_U} - 1) \times T}{2^{Level_U}} \right] \right\rceil$.

When we search for a record, we first search for the record node containing the record and then search within the record node to find the record. For a record node at an upper level, we just need to access the upper levels from the root to reach it. For a record node in a lower level, we must traverse (access) all the upper levels first and then search the lower levels. The search process in the lower levels depends on the specific implementation: for software T-tree, using either CMOS or memristors, we search the corresponding subtree with the search process similar to the search process in the upper levels; for Hash-CAM and T-tree-CAM, we search the corresponding CAM partition with the comparison and match signal combination described in Section 4; for TB$^+$-tree, we go down the B$^+$-tree to the leaf level and search the internal nodes with the comparison and match signal combination described in Section 4 to decide which subtree has the input key; for TB$^+$-tree-CAM, we go through a per-defined number of B$^+$-tree levels and search all the leaves in the corresponding subtree with the comparison and match signal combination described in Section 4.

The time required to reach a node at the $i^{th}$ upper level is $i \times NodeTime_U$ where $i \in Z$ and $i \in [1, Level_U]$. The time required to reach a node at the $i^{th}$ lower level in the T-tree is $Level_U \times NodeTime_U + i \times NodeTime_{LT}$ where $i \in Z$ and $i \in [1, Level_{LT}]$. The time required to reach a node at the $i^{th}$ lower level in the TB$^+$-tree is $Level_U \times NodeTime_U + Level_{LTB} \times NodeTime_{LTB}$ where $i \in Z$ and $i \in [1, Level_{LTB}]$.

---

$AvgTime_T =$
$\{NodeTime_U \times [(Level_U - 1) \times 2^{Level_U} + 1]$
$+ NodeTime_U \times Level_U \times (N_T - 2^{Level_U} + 1)$
$+ NodeTime_{LT} \times [(Level_{LT} - 2) \times 2^{(Level_U + Level_{LT} - 1)} + 2^{Level_U}]$
$+ NodeTime_{LT} \times Level_{LT} \times [N_T - (2^{(Level_U + Level_{LT} - 1)} - 1)]\} / N_T$

$AvgTime_{TB} = \{NodeTime_U \times [(Level_U - 1) \times 2^{Level_U} + 1]$
$+ NodeTime_U \times Level_U \times (N_{TB} - 2^{Level_U} + 1)$
$+ NodeTime_{LTB} \times Level_{LTB} \times (N_{TB} - 2^{Level_U} + 1)\} / N_{TB}$

$AvgTime_{HC} = HashTime + MemCAMSeachTime$

$AvgTime_{TC} = NodeTime_U \times Level_U + MemCAMSeachTime$

$AvgTime_{TBC} = NodeTime_U \times Level_U + NodeTime_{LTB} \times Level_{LTB}$
$+ MemCAMSeachTime$

**Figure 14: Access Time Equations**

---

**Table 4: Records For Each Data Structure**

| Data structure | Number of records |
|---|---|
| Software CMOS-based T-tree | 5.4 x 10$^9$ |
| Memristor-based T-tree | 3.4 x 10$^{11}$ |
| Hash-CAM | 6.9 x 10$^{10}$ |
| T-tree-CAM | |
| TB$^+$-tree | 2.8 x 10$^{10}$ |
| TB$^+$-tree-CAM | |

We assume a random uniform distribution of keys to search and define the average record search time as the average time required to reach the corresponding record node. We also assume that each record node has the same number of records and calculate the average record search time by dividing the sum of the time required to reach each record node by the number of record nodes.

Figure 14 shows how we calculate the average record search time for each of the five memristor-based data structures. We calculate the average record search time of a CMOS-based T-tree and a memristor-based T-tree by changing the value of $NodeTime_{LT}$ in $AvgTime_T$. $AvgTime_{TB}$ is the average record search time of a hybrid CMOS-memristor TB$^+$-tree. We also calculate the average record search time of Hash-CAM ($AvgTime_{HC}$), T-tree-CAM ($AvgTime_{TC}$), and TB$^+$-tree-CAM ($AvgTime_{TBC}$) based on the MemCAM search time described in Section 4. The average record search time of Hash-CAM is the sum of the time to access the hash table and the time to search a CAM partition. The average record search time of a T-tree-CAM is the sum of the time to go through all upper levels in the T-tree and the time to search a CAM partition. The average record search time of TB$^+$-tree-CAM is the sum of the time to go through all upper levels in the T-tree, the time to go through a predefined number of levels in a B$^+$-tree, and the time to perform CAM-like search operation in a subtree. We do not consider the records in upper levels when we calculate the average record search time for these three hybrid storage structures because most of the records are in lower levels.

## 6.3 Modeling Results and Analysis

Our evaluation is based on 32MB cache and 128GB DRAM [33]. Based on the densities of DRAM (15Gbit/cm$^2$ [3]) and a memristor array (1Tbit/cm$^2$ [32]), we assume an 8TB memristor memory and a 1TB MemCAM. We choose the number of records in a T-node (T) to be 10 and store 17 levels of T-tree in cache. We choose the number of records in a B$^+$-tree node (B) to be 80 to balance between the depth and node utilization rate of the B$^+$-tree. Table 4 shows the number of records we can store in different data structures assuming 100% node utilization rate for T-tree and 75% node utilization rate for B$^+$-tree.

For CMOS-related parameters, we use the data from performance analysis of Intel's latest processors [21] for $NodeTimeU$ (16ns) and $NodeTimeLC$ (60ns). For memristor-related paramters, we can still achieve 1-ns switching time as the scale of memristor goes from 50nm to 5nm (density going from 10Gbit/cm$^2$ to 1TGbit/cm$^2$) since memristor switching time is proportional to $R_{OFF}/R_{ON}$. The RC delay of wires increases as memristors scale down but we can use technologies such as repeater insertion in order to achieve the same RC delay at 5-nm scale and 50-nm scale. As a result, we can achieve similar search time at both scales. It requires one write (to write the key), two reads (one to read the comparison results, and one to read the address of the next node), and a comparison to search a node in the $TB^+$-tree. We use 8-byte keys and 8-byte pointers thus the comparison time is 136ns. We call the time required to perform a read/write operation to the memristor array from the peripheral circuitry memristor arrayread/write latency. We need peripheral circuitry to decode the address, evaluate the data read out (for reads) and decide the applied voltages based on the data to write (for writes) when we access the memristor array externally. As a result, the memristor array read/write latency (which varies from 10ns [22, 29] to 120ns [28]) is much longer than the step time of 2ns (where the operations occur inside the memristor array and the applied voltages are known beforehand). As a result, *NodeTimeLM* varies from 166ns to 496ns.

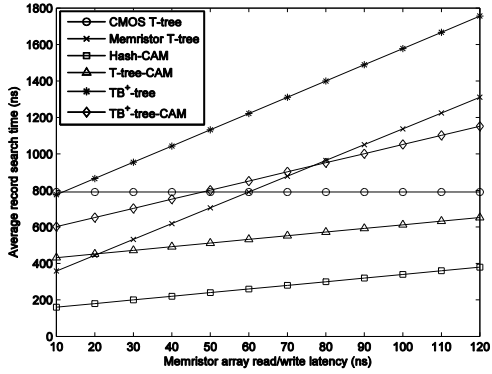

**Figure 15: Performance of various data structures vs. memristor latency (number or records, R, = $10^9$)**

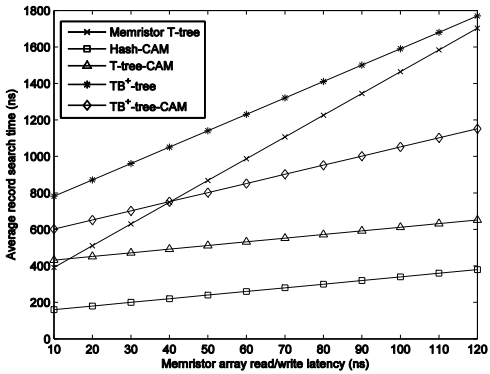

**Figure 16: Performance of various data structures vs. memristor latency (number or records, R, = $10^{10}$)**

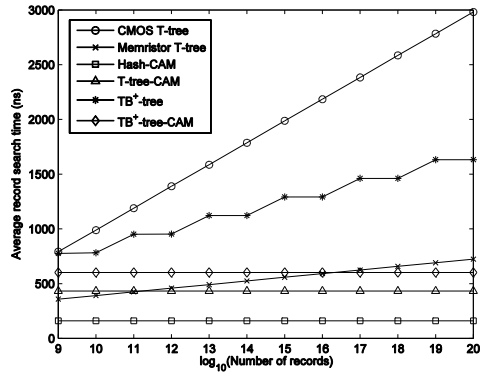

**Figure 17: Search Time vs. number of records**

We first choose the number of records (R) to be $10^9$, therefore all the records can be stored in any of the six data structures, and compare the performance of the six data structures shown in Table 4. Figure 15 shows the average record search time as we increase memristor array read/write latency from 10ns to 120ns. We can see that only Hash-CAM and T-tree-CAM perform better than memory-based T-trees. The performance of $TB^+$-tree-CAM is between the performance of CMOS T-tree and Memristor T-tree and the performance of $TB^+$-tree is worse than both memory-based T-trees.

We then choose the number of records (R) to be $10^{10}$, which means that the capacity of DRAM is no longer high enough to store all the records. Figure 16 shows the average record search time as we increase memristor array read/write latency from 10ns to 120ns. We do not show the performance of software/CMOS + DRAM in Figure 16 since it requires disk accesses for this data size and the average record search time increases to milliseconds. We can see from Figure 16 that Hash-CAM and T-tree-CAM still perform better than memristor-based T-tree. $TB^+$-tree still performs worse than memristor-based T-tree but the performance gap decreases as memristor array read/write latency increases. $TB^+$-tree-CAM outperforms memristor-based T-tree as long as memristor array read/write latency is longer than 40ns. In general, $TB^+$-tree and $TB^+$-tree-CAM perform better than when R is $10^9$. The reason is that when R is $10^9$, the node utilization rate of $B^+$-trees is low, which means that we cannot benefit from the shallowness of $B^+$-trees.

We then change the number of records from $10^9$ to $10^{20}$ to see how data size affects performance of the six data structures. From Figure 16 we can see that memristor-based T-tree has better performance when memristor array read/write latency is lower, so we choose memristor array read/write latency to be 10ns to have the best possible performance of memristor-based T-tree. The results are shown in Figure 17. We can see that the search time of CMOS-based T-tree, memristor-based T-tree, and $TB^+$-tree increase as the number of records increases. The search time of Hash-CAM, T-tree-CAM, and $TB^+$-tree-CAM remain almost the same. When the number of records goes beyond $10^{16}$, $TB^+$-tree-CAM outperforms memristor-based T-tree.

We also calculate the theoretical maximum lifetime of the four hybrid data structures assuming continuous search operations, which is shown in Table 5. Generally, lifetime increases as search time increases, since increased search time results in reduced write frequency. The exception here is T-tree-CAM, whose lifetime is limited by the capacity of cache. However, we can see that even in the worst case we can achieve a one-year lifetime. If we take into account the time required to read out the matched records, the lifetime will be even longer.

From the above analysis, we observe that reducing memristor external read/write latency below 40ns and if the number of records is smaller than $10^{16}$ then memristor-based T-tree is the best data structure for search applications. Otherwise, $TB^+$-tree-CAM performs better through combined compute and storage and has an acceptable lifetime.

## 7. Related Work

A memristor-based crossbar memory system has been demonstrated by HP Labs [35]. Strategies and peripheral circuitry have been designed to write into and read from a memristor array. The memristor memory demonstrates much higher density and access time comparable to CMOS RAMs. A hybrid CMOS-memristor CAM has also been proposed to achieve larger capacity [11]. However, combining CMOS transistors with memristors reduces bit density and increases manufacturing difficulty.

Recently several researchers have explored spin-based devices for both storage and computing [15, 16, 27]. In STT-MRAM [16], one CMOS transitor and one magnetic tunnel junction (MTJ) are combined to build a cell. In MTJ-based logic units [26], MTJs are used to perform logic operations on data stored in other MTJs. Our work focuses on memristor technology with wear out constraints. Exploring more general configurable accelerators with different technologies is an interesting area of future work.

Other emerging memory technologies with endurance problems include PCM (Phase Changing Memory). Recent work proposes new B+-tree algorithms for PCM to improve performance and reduce writes [7]. They design unsorted node organizations to reduce the number of writes incurred during insert and delete operations. Their results show that an approach where only the leaves are unsorted performs better than when all the nodes are unsorted. The reason is that it requires linear time to search within a node in PCM. However, with the computation ability of memristors, we can perform simultaneous comparisons and reduce search time within a node, which makes the unsorted scheme a better choice for in-place computing technologies.

Extensive research has been performed on processing-in-memory (PIM) to improve performance by combining processing units and memory [10, 14, 24, 26]. Terasys [14] augments a standard 4-bit memory with a single-bit ALU controlling each column of memory. In DAAM (Dynamic Associative Access Memory) [24] a large number of small processing elements are put in a DRAM's sense amps. DIVA [10] incorporates multiple PIM chips to a conventional microprocessor. Smart Memories [26] has multiple processing tiles which can be configured based on the requirements of applications. The main idea of PIM is to combine compute and storage, similar to our proposed data structures.

Recent research [23] "disaggregated" memory to expand and share memory across servers. With memory capacity increase, we are able to store more data and obtain more benefits from our hybrid data structures. Furthermore, by performing in-place computation the disaggregated memory could serve as an application appliance rather than simply memory.

## 8. Conclusion

Memristors are an emerging technology with potential to provide high-density storage augmented with in-place computing through implication logic. In this paper we explore this combined storage compute as a method to accelerate point and range search queries, which serve as specific instances of more general configurable accelerators. We first show how to use implication logic to create a configurable CAM that can support both point and range queries; however, low endurance of memristors limits the benefit we can obtain from these storage structures. To more fully utilize the computation ability of memristors and overcome the endurance problem, we introduce novel data structures for use with memristor-based storage+compute structures.

**Table 5: Theoretical maximum lifetime of four hybrid data structures ($T_{access}$: memristor array read/write latency)**

| Data structure | Lifetime (years) | | |
|---|---|---|---|
| | $T_{access}$=10ns | $T_{access}$=60ns | $T_{access}$=120ns |
| Hash-CAM | 2.4 | 3.9 | 5.7 |
| T-tree-CAM | 0.8 | 1.0 | 1.2 |
| $TB^+$-tree | 88.6 | 139.5 | 200.7 |
| $TB^+$-tree-CAM | 68.2 | 96.6 | 130.6 |

We first propose MemCAM, a configurable memristor-based CAM design. The computation ability of memristors makes it possible to perform range search using MemCAM while the high density of memristors provides an opportunity to build CAMs with large capacity and small area. We use SPICE to model memristor power and performance. With 50nmx50nm memristors and a K-bit search word, for a MemCAM supporting both point and range queries, the energy consumption is $(0.44+0.82*\log_2(K))$ fJ/bit/search and the search time is $(16+20*\log_2(K))$ ns, and for MemTCAM supporting both point and range queries, the energy consumption is $(0.83+0.82*\log_2(K))$ fJ/bit/search and the search time is $(22+20*\log_2(K))$ ns.

We then propose a series of configurable hybrid data structures using both conventional CMOS cache hierarchies and memristor technologies to solve the endurance problem. These data structures can be reconfigured to trade between performance and lifetime and to adapt to future memristors with improved endurance. We use an analytic model to calculate and compare the performance and lifetime of two memory-based T-trees and four hybrid data structures. The results show that hybrid data structures can utilize MemCAM search abilities and improve lifetime from seconds to years. Furthermore, $TB^+$-tree-CAM, a hybrid CMOS-memristor data structure combining T-tree, $B^+$-tree and CAM, manages to balance between performance and lifetime and can outperform other data structures when taking both performance and lifetime into consideration.

## 9. Acknowledgements
Thanks to those that funded and contributed.